\documentclass{jfm}

\usepackage{graphicx}
\usepackage{natbib}
\usepackage{xcolor}
\usepackage{color}

\ifCUPmtlplainloaded \else
  \checkfont{eurm10}
  \iffontfound
    \IfFileExists{upmath.sty}
      {\typeout{^^JFound AMS Euler Roman fonts on the system,
                   using the 'upmath' package.^^J}%
       \usepackage{upmath}}
      {\typeout{^^JFound AMS Euler Roman fonts on the system, but you
                   dont seem to have the}%
       \typeout{'upmath' package installed. JFM.cls can take advantage
                 of these fonts,^^Jif you use 'upmath' package.^^J}%
      }
  \else
  \fi
\fi

\ifCUPmtlplainloaded \else
  \checkfont{msam10}
  \iffontfound
    \IfFileExists{amssymb.sty}
      {\typeout{^^JFound AMS Symbol fonts on the system, using the
                'amssymb' package.^^J}%
       \usepackage{amssymb}%

      }{}
  \fi
\fi

\ifCUPmtlplainloaded \else
  \IfFileExists{amsbsy.sty}
    {\typeout{^^JFound the 'amsbsy' package on the system, using it.^^J}%
     \usepackage{amsbsy}}
    {\providecommand\boldsymbol[1]{\mbox{\boldmath $##1$}}}
\fi

\newsavebox{\astrutbox}
\sbox{\astrutbox}{\rule[-5pt]{0pt}{20pt}}

\usepackage{siunitx}
\usepackage{tabularx}

\title[Temperature and velocity Lagrangian measurements in turbulent convection]{Simultaneous temperature and velocity Lagrangian measurements in turbulent thermal convection}

\author[O. Liot \emph{et al.}]%
{O. Liot,$^{1}$
F. Seychelles,$^1$
F. Zonta,$^2$
S. Chibbaro,$^{3,4}$
T. Coudarchet,$^1$\break
Y. Gasteuil,$^1$
J.-F. Pinton,$^1$
J. Salort,$^1$
and F. Chill\`a,$^1$\thanks{Email address for correspondence: francesca.chilla@ens-lyon.fr}}

\affiliation{$^1$Laboratoire de Physique, ENS de Lyon, 46 all\'ee d'Italie, 69364 Lyon Cedex 7, France, EU\\[\affilskip]
$^2$Dip. Ing. Elettrica, Gestionale e Meccanica, Via delle scienze 208, 33100 Udine, Italy, EU\\[\affilskip]
$^3$Sorbonne Universit\'es, UPMC Univ Paris 06, UMR 7190, Institut Jean Le Rond d'Alembert, F-75005, Paris, France\\
$^4$CNRS, UMR 7190, Institut Jean Le Rond d'Alembert, F-75005, Paris, France}

\pubyear{2015}
\volume{650}
\pagerange{119--126}
\date{05 June 2015; revised ?; accepted ?. - To be entered by editorial office}
\graphicspath{{./}{./skewness/}{./PDFT_conv/}{./PDFNu_puissances/}{./Spectres-vx/}{./Spectres-vz/}{./Spectres-T/}{./Comparaison/}{./Traj-NuLz-ok/}{./Traj-Ture/}{./map-T/}{./Vfluct/}}
\usepackage{graphics}%
\begin{document}

\maketitle

\begin{abstract}
We report joint Lagrangian velocity and temperature measurements in turbulent thermal convection. 
{
Measurements are performed using an improved version (extended autonomy) of the neutrally-buoyant instrumented particle \citep{RSI} that was used by 
\cite{gasteuil} to performed experiments in a parallelepipedic Rayleigh-B\'enard cell.
}
The temperature signal is obtained from a RF-transmitter. Simultaneously, we determine particle's position and velocity with one camera, which grants access to the Lagrangian heat flux. 
{ Due to the extended autonomy of the present particle, we obtain well converged temperature and velocity statistics, as well as 
pseudo-eulerian maps of velocity and heat flux.
Present experimental results have also been compared with the results obtained by a corresponding campaign of Direct Numerical Simulations and Lagrangian Tracking of massless tracers.
The comparison between experimental and numerical results show the accuracy and reliability of our experimental measurements.
Finally, the analysis of lagrangian velocity and temperature frequency spectra is shown and discussed.
In particular, we observe that temperature spectra exhibit an anomalous $f^{-2.5}$ frequency scaling, likely representing the ubiquitous passive and active scalar behavior of temperature.
}
\end{abstract}

\begin{keywords}
{You can use up to three keywords, but you have to choose them from a list of possible alternatives when you submit the paper}
Rayleigh-B\'enard convection, large-scale circulation, \emph{smart particle}, turbulent convection, Lagrangian measurements.
\end{keywords}
%
%

\section{Introduction}

 Thermal convection occurs in many industrial and geophysical applications, ranging from heat exchangers or nuclear/chemical reactors to atmspheric {\bf circulation}.
In most of these cases, the flow is highly turbulent and transports heat very efficiently.
Nevertheless, understanding \textcolor{red}{and modeling} the local and global properties of the temperature, velocity and heat-flux fields {\bf in these situations} is still a challenge. To {\bf analyze} turbulent thermal convection in a laboratory, we choose the Rayleigh-B\'enard configuration: a horizontal layer of fluid confined between a cooling plate above and a heating plate below. 
In this flow  configuration, the temperature gradient is confined in the thermal boundary layers close to the heating and cooling plates. 
{\bf Due to the strong mixing, a nearly homogeneous temperature distribution is observed in the bulk.
The driving force of the flow is measured by the Rayleigh number:}

\begin{equation}
Ra = \frac{g\alpha \Delta T H^3}{\nu \kappa},
\end{equation}

where $H$ is the height of the cell, $g$ is the gravitational acceleration, $\alpha$ is the constant pressure thermal expansion coefficient, $\Delta T=T_h - T_c$ is the difference of temperature between the heating and the cooling plate, $\nu$ is the kinematic viscosity of the fluid, and $\kappa$ is its thermal diffusivity. The Prandtl number {\bf expresses the ratio between viscous and thermal dissipation}:

\begin{equation}
Pr=\frac{\nu}{\kappa}. 
\end{equation}

A further input parameter is the aspect ratio $\Gamma$, {\bf i.e.} the ratio between the horizontal and the vertical size of the cell.
The response of the system {\bf is} represented by the Nusselt number, {\bf which compares convective and conductive}  heat flux:

\begin{equation}
Nu = \frac{QH}{\lambda \Delta T},
\end{equation} 

where $Q$ is the global heat flux, and $\lambda$ is the thermal conductivity of the fluid. 

Assuming locally homogeneous and isotropic turbulence, for sufficiently high Reynolds and Peclet numbers, passive scalars and velocity spectra follow the well-known Kolmogorov-Obukhov laws \citep{kolmogorov,Monin}.
However, in thermally driven flow, temperature is not a passive scalar and the similarity theory {\bf requires a further generalization}. It has been {\bf argued} that for small scales a thermally stratified fluid can be considered locally stationary and
homogeneous, but not isotropic and axially symmetric relative to the vertical direction \citep[]{Bolgiano,Obukhov59}.
{\bf Within this framework}, the scalings become dependent on the Bolgiano-Obukhov (BO59) lengthscale measuring the importance of the thermal stratification,
%
%
\begin{equation}
L_B \equiv \epsilon_u^{5/4} \epsilon_T^{-3/4} \left(\alpha g\right)^{-3/2},
\end{equation}
where $\epsilon_T$ and $\epsilon_u$ are the temperature and kinetic energy dissipation rate, respectively.
 $L_B$ characterizes the minimum length scale of inhomogeneities beyond which stratification should {\bf be} taken into account.
The ordinary Kolmogorov spectrum scalings ($k^{-5/3}$) are expected to be recovered at scales smaller than the Bolgiano length.
On the other hand, if $L_B$ becomes much larger than the external turbulence length scale $L_0$, the effect of the mean flow becomes important and the similarity theory does not apply. {\bf Yet}, all the {\bf theoretical predictions and} scalings are {\bf assessed} for regions of space far enough from
solid boundaries.
In general, stratification affects a certain range of scales, which cannot be considered locally isotropic.
Corrections to the velocity and temperature correlations and spectra may be universal and may be in principle
determined empirically. 
{For stratified flows, \bf \cite{Bolgiano} established a theoretical framework to determine the asymptotic form of these functions for scales much greater than $L_B$.}
%
These Bolgiano-Obukhov (BO59) scalings predict spectra {\bf that are} steeper for the velocity and {\bf milder} for the temperature compared to those given by Kolmogorov-Obukhov scalings.
Nevertheless, the ultimate picture on velocity and temperature scalings in thermal convection is far from being obtained \citep{review3}.
%
Experimental measurements at $Ra \approx 10^{10}$ and $Pr = 4.4$ exhibit a K41 behaviour for velocity \citep{zhou} and for velocity and temperature \citep{sun} structure functions. However, direct numerical simulations (DNS) of thermal convection on a cilindrical domain \citep{kunnen} show a BO59 scaling for temperature structure functions in the radial direction and for vertical velocity structure functions in the longitudinal direction ($Ra =10^{8}$, $Pr = 1$). 

Even though remarkable progresses have been  made towards a deeper understanding of  scaling laws between control and response parameters 
\citep{GLmodel,chavanne,stevens2013}, new investigations are required to clarify some of the open issues  \citep{review3}.
Most of literature studies on Rayleigh-B\'enard turbulence have been focused on eulerian measurements of velocity and temperature distribution \citep{xia,tilgner}, with the aim of characterizing the behavior of the local heat flux, $Nu$, 
as a function of the Rayleigh number, $Ra$ \citep{shang}.
{\bf Although the mean velocity in the central region of the convection cell is homogeneous and close to zero, the velocity root-mean-square is neither homogeneous nor isotropic \citep{xia,qiu2000}.
These flow inhomogeneities prevent from using the frozen-flow hypothesis \citep{taylor1938}: this makes the connection between time-domain measurements and space-domain predictions difficult.}
%

More recently, improvements in computing power and storage capacities has allowed the appearance of lagrangian studies of turbulence, {\bf which naturally provide useful informations on transport mechanisms}.
The Lagrangian description of turbulence {\bf has significantly contributed to our current comprehension of transfer processes} \citep{reviewLag}.
In particular, a number of experimental \citep{mordant,porta,voth} and numerical \citep{yeung2002,biferale2004} studies focused on velocity and acceleration statistics in homogeneous and isotropic turbulence.
The first numerical lagrangian studies of thermal convection {\bf were} those of \cite{schumi1,schumi2}, which were specifically focused on 
pair dispersion and on acceleration statistics.
Although the tracer motion {\bf was} largely anisotropic due to the vertical buoyancy, pair dispersion {\bf was} close to the homogeneous and isotropic turbulence. Acceleration and temperature statistics showed a non-gaussian behaviour characterized by a large intermittency (higher in the horizontal directions).
Relevant to the present work {\bf was} also the finding of the non-symmetric  behavior of the heat transport probability density function.
From an experimental point of view, measurements of turbulent thermal convection in a Lagrangian framework are relatively scarce.
Only recently, \cite{xia_ptv}  used three-dimensional particle tracking velocimetry to analyze velocity and acceleration statistics in turbulent Rayleigh-B\'enard convection. 
{\bf In particular, they observed a gaussian and a stretched exponential distribution for the probability density function of velocity and acceleration in the centre of the cell.}
%

From the above review, phenomenological and statistical analyses of turbulent Rayleigh-B\'enard convection in a Lagrangian frame of 
reference appear not yet complete and require further investigation. This is exactly the purpose of the present study.
In this work, we present  both experimental and numerical measurements of temperature and velocity in a Lagrangian frame.
For the experiments, we improved the neutrally-buoyant instrumented particle presented in \cite{RSI} and already 
used and tested by \cite{gasteuil}.
This \emph{smart particle} explores a rectangular Rayleigh-B\'enard cell filled with water. We compare Eulerian maps obtained from our Lagrangian data with PIV measurements, to show that the particle samples correctly the entire flow (and to deduce pseudo-Eulerian maps of temperature and thermal flux).
{\bf At the same time, we perform Direct Numerical Simulations (DNS) and Lagrangian particle tracking of massless tracers in 
turbulent Rayleigh-B\'enard convection.}
These numerical simulations correspond to the ideal case of a particle of zero diameter and, therefore, they can be viewed as a "conceptual experiment" built to highlight possible finite-size effect of the smart-particle in the experiments. 
Velocity, temperature and heat flux statistics obtained from experiments and numerical simulations are compared and discussed.

%

\section{\emph{Smart particle}, experimental setup and numerical simulation method}

 \begin{figure}
\begin{minipage}{0.54\textwidth}
\begin{center}
(a)
\includegraphics[width=1\textwidth]{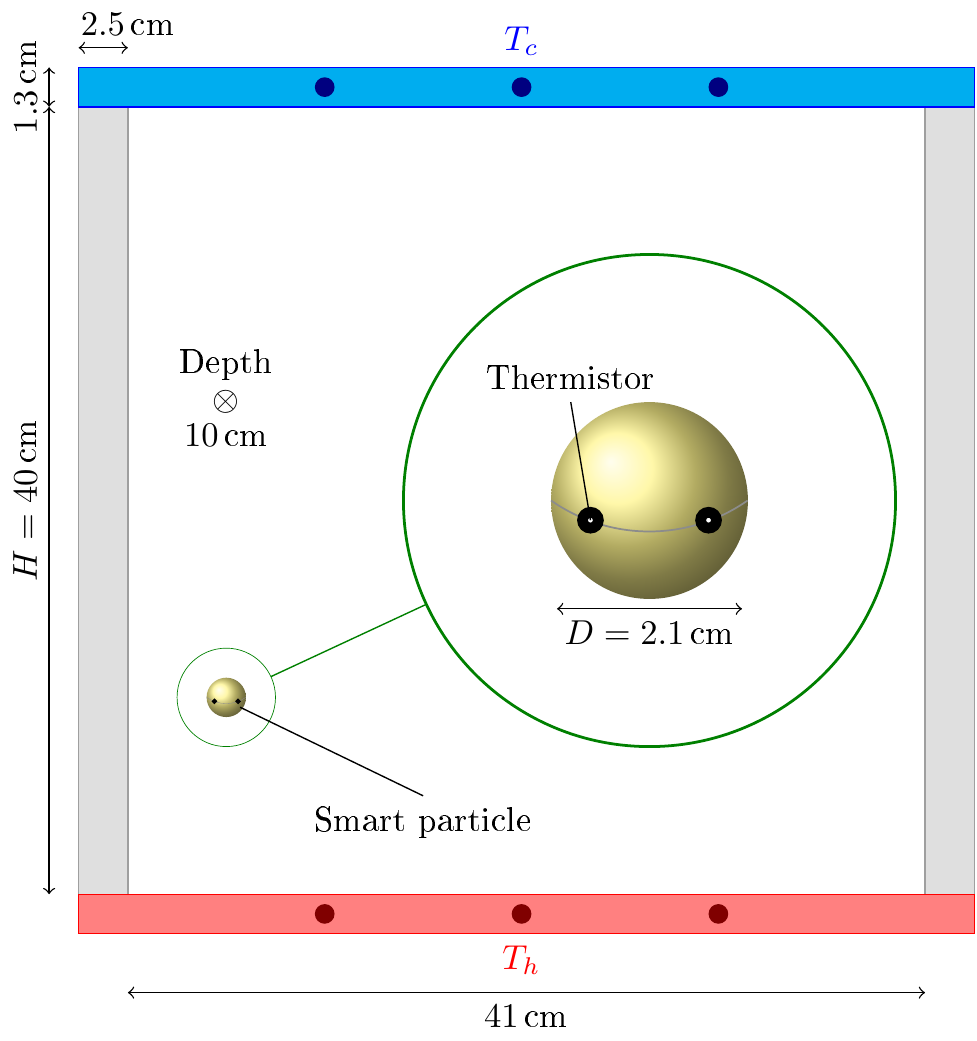}
\end{center}
\end{minipage}
\begin{minipage}{0.02\textwidth}
~~
\end{minipage}
\begin{minipage}{0.44\textwidth}
\begin{center}
(b)

\vspace{0.2cm}

\includegraphics[width=0.9\textwidth]{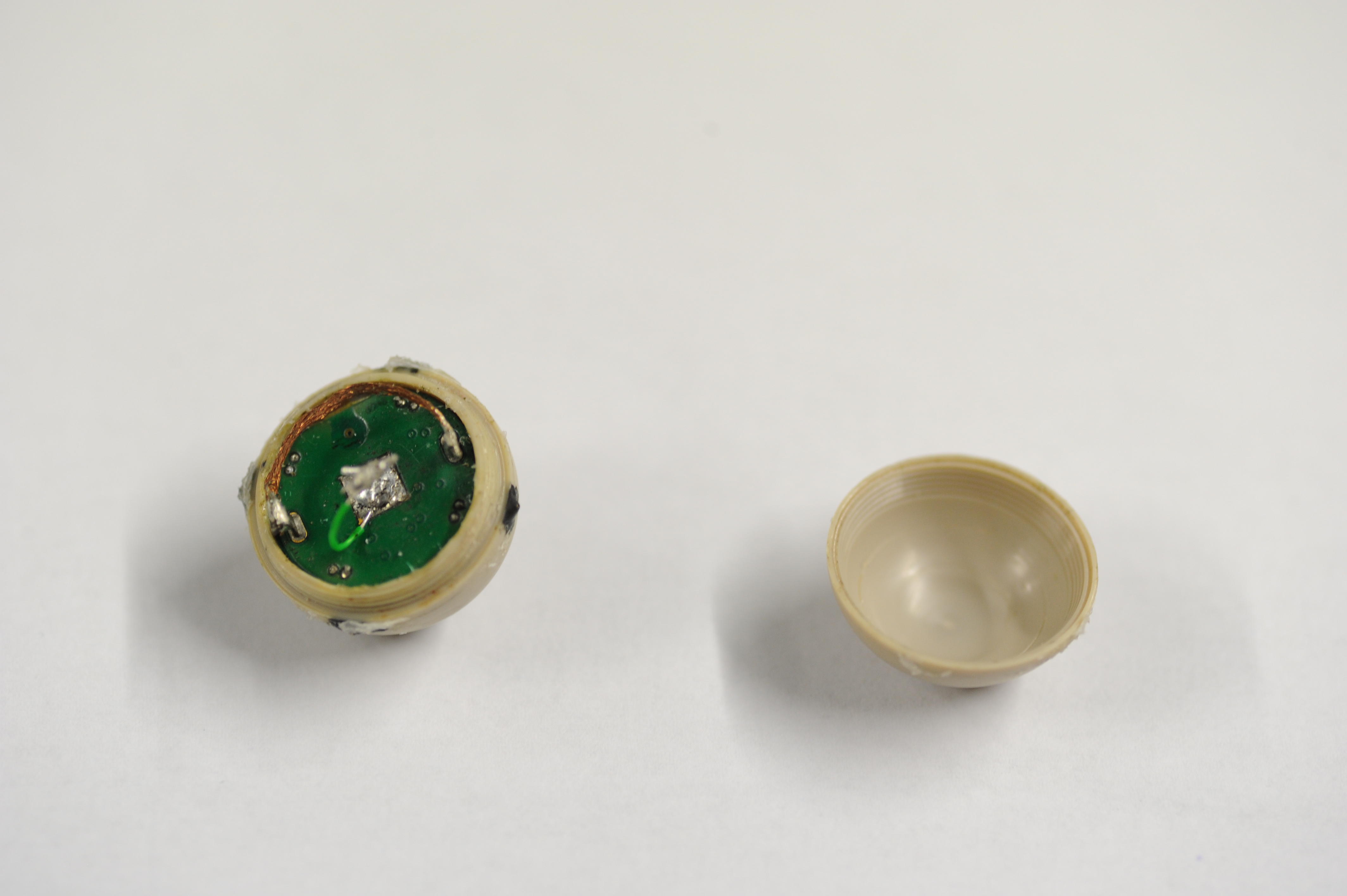}
\end{center}
   \end{minipage}
   \caption{(a) Sketch of the convection cell and of the mobile sensor. The six dark dots in the plates show the location of the PT100 temperature sensors. (b) Photograph of the instrumented particle (open). We see the batteries place.}
   \label{cell}
   \end{figure}

\subsection{\emph{Smart Particle}}

The mobile sensor consists in a 2.1 \si{cm} in diameter capsule containing temperature instrumentation, a radio-frequency emitter, and two batteries. Four cylindrical thermistors (0.8 mm in length, 0.4 mm in diameter, 230 $k\Omega$, response time 0.6 s in water) are mounted on the capsule wall protruding 0.5 mm into the surrounding flow (see sketch in zoom of figure \ref{cell}(a) and photograph figure \ref{cell}(b)). A resistance controlled oscillator is used to create a square wave whose frequency depends on the average of the four measured temperatures. This square wave is used directly to modulate the frequency of a radio wave generated by the radio-frequency emitter. The temperature signal is recovered on the fly by a stationary receiver. The capsule has been redesigned (compared to the one described by \cite{RSI}) in order to have the four thermistors at the equator and a simpler handling. A new shell has been conceived to  {\bf contain} two batteries and to extend the emission time which can now reach up to 1000 turnover times. At the same time the trajectory of the particle is recorded with a digital camera placed in front of the large face of the cell (due to cell dimensions, we assume that the mean flow is quasi bi-dimensional).

\subsection{Experimental setup}

Our convection cell is a 10.5 \si{cm}-thick 41.5 \si{cm} $\times$ 41.5 \si{cm} rectangular cell with 2.5 \si{cm}-thick PMMA walls (see sketch figure \ref{cell}(a)). Both plates consist in 4 \si{cm}-thick copper plates coated with a thin layer of nickel. The bottom plate is Joule-heated while the top plate is cooled with a temperature regulated water circulation. Plate temperatures are controlled by PT 100 temperature sensors. We work with deionized water. The bulk temperature is fixed between 37.05\si{\degreeCelsius} and 38.35\si{\degreeCelsius} in different experiments. The corresponding Prandtl number are in the range $4.62-4.49$. Main parameters are grouped in table \ref{table:tableau}.

\begin{table}
\begin{center}
\begin{tabular}{ccc}
$\Delta T (\si{\degreeCelsius})$ & $Ra$ & $Nu$\\
\hline 
13.15 & $3.5\times 10^{10}$ & 230\\ 
18.60 & $5.0\times 10^{10}$ & 244\\ 
22.90 & $6.2\times 10^{10}$ & 264\\
\end{tabular}
\end{center}
\caption{Parameters used for acquisitions in the experiments.}
\label{table:tableau}
\end{table}

\subsection{Numerical simulations}

\begin{figure}
\begin{center}
\includegraphics[width=0.80\textwidth]{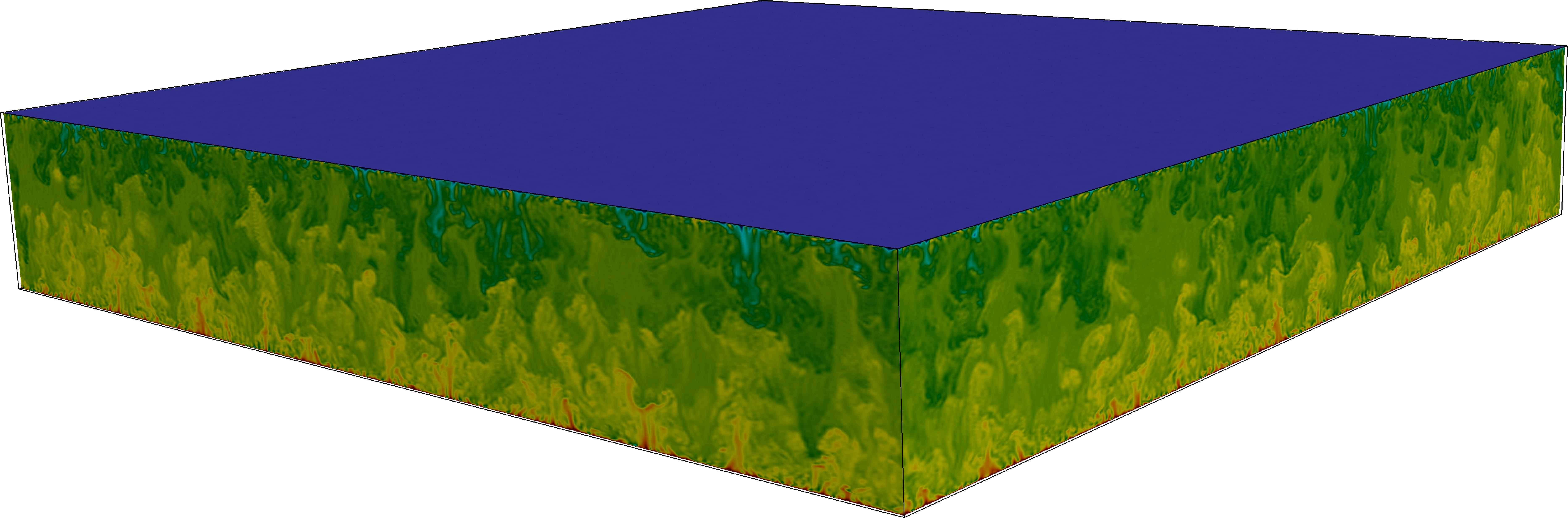}
\caption{Sketch of the numerical simulation domain. Temperature contours for $Ra=10^9$ are also shown (red indicates regions of high temperature whereas blue indicates regions of low temperature). }
\label{sketch}
   \end{center}
   \end{figure}

Direct Numerical Simulations are performed to complement our experimental results.
We consider an incompressible and Newtonian turbulent flow of water confined between two rigid boundaries. Streamwise, spanwise and wall-normal coordinates are indicated by $x$, $y$ and $z$, respectively.
The bottom wall is kept at uniform high temperature ($T_{h}$), whereas the top wall is kept at uniform low temperature ($T_{c}$). 
The size of the computational domain is $L_x \times L_y \times L_z = 4\pi H \times 4\pi H \times 2H$ (in $x$, $y$ and $z$, respectively), where $H$ is the half-channel height.
A sketch of the computational domain/flow conditions is presented in Fig.\ref{sketch}.

The imposed temperature difference $\Delta T=\left( T_{h}-T_{c}\right)$ 
 between  the bottom and the top wall 
induces an unstable buoyancy effect within the flow field 
(the acceleration due to gravity $g$ acts downward along $z$).
Mass, momentum and energy equations in dimensionless form and under the Boussinesq approximation are:
\begin{equation}
\nabla \cdot {\bf u} = 0, 
\label{c1}
\end{equation}
\begin{equation}
\frac{\partial {\bf u}}{\partial t} + ( {\bf u}\cdot{\nabla  ) {\bf u}} = -\nabla p + 4\sqrt{\frac{Pr}{Ra}}\nabla^2 {\bf u}-\delta_{i,3} \theta, 
\label{ns1}
\end{equation}
\begin{equation}
\frac{\partial {\bf \theta}}{\partial t} + ({\bf u}\cdot{\nabla ) { \theta}} = + \frac{4}{\sqrt{Pr Ra}}\nabla^2 {\theta}, 
\label{en1}
\end{equation}
where $u_i$ is the $i^{th}$ component of the velocity vector, $\theta$ is the dimensionless temperature $\theta= (T-T_{ref})/\Delta T$, $p$ is pressure, whereas $\delta_{i,3} \theta$ is the buoyancy force (acting in the vertical direction only) that drives the flow.
Eqs. \ref{c1}-\ref{en1} have been obtained using $h$ as reference length, $u_{ref}=\sqrt{g\alpha \Delta T/2 H}$ as reference velocity, $T_{ref}=(T_h+T_c)/2$ as reference temperature and $p=\rho g \alpha H \Delta T /2$ as reference pressure. Density $\rho$, kinematic viscosity $\nu$, thermal diffusivity $\kappa$ and thermal expansion coefficient $\alpha$ are evaluated at a mean fluid temperature of $\simeq 30 ^o C$.  
The Prandtl and the Rayleigh numbers in Eqs. \ref{c1}-\ref{en1} are defined as $Pr=\nu/k$ and $Ra= (g \alpha \Delta T (2H)^3) / (\nu k)$, respectively. 
In the present study, we keep the Prandtl number $Pr=4$ and we vary the Rayleigh number between $Ra=10^7$ and $Ra=10^9$.
Periodic boundary conditions are imposed on velocity and temperature along the streamwise $x$ and spanwise $y$ directions; at the walls, no slip conditions are enforced for the momentum equations while constant temperature conditions are adopted for the energy equation. 
The resulting set of equations are discretized using a pseudo-spectral method based on transforming the field variables into wavenumber space, through Fourier representations for the periodic (homogeneous) directions $x$ and $y$, and Chebychev representation for the wall-normal (non-homogeneous) direction $z$ (see \cite{zonta2012,zonta2014} for details).
We used up to $512 \times 512 \times 513$ grid points to discretize the computational domain.
We injected $N_p=1.28 \cdot 10^5$ lagrangian tracers and we computed their dynamics as
\begin{equation}
\dot{\bf x}_p = {\bf u}\left( {\bf x_p}(t), t\right)~~~~~
{ \theta_p} = {\theta}\left( {\bf x_p}(t), t\right), 
\label{lag}
\end{equation} 
with ${\bf x_p}$ the tracers position and $\theta_p$ their temperature.
Velocity and temperature at particle position are obtained by $6^{th}$ order Lagrange polynomials.  
Time advancement for the Lagrangian tracers is achieved using a $4^{th}$ order Runge-Kutta scheme.

As discussed above, numerical simulations are carried out in a laterally-opened (along $x$ and $y$) domain configuration, which is similar to that considered in the theory 
(where the vertical direction is not homogeneous, because of the buoyancy, while the other two are homogeneous
and isotropic). 
Experiments must be performed in a \textcolor{blue}{closed} cell and some anisotropic effects may appear also 
in the other directions. 
Moreover, passive tracers in numerical computations are free to explore the entire domain,
whereas the smart-particle cannot access the small region near-to-the-wall, given its dimensions.
These differences might have some effects, as will be discussed in the following.

\section{Lagrangian measurements}
\begin{figure}
\begin{center}
\includegraphics[width=0.45\textwidth]{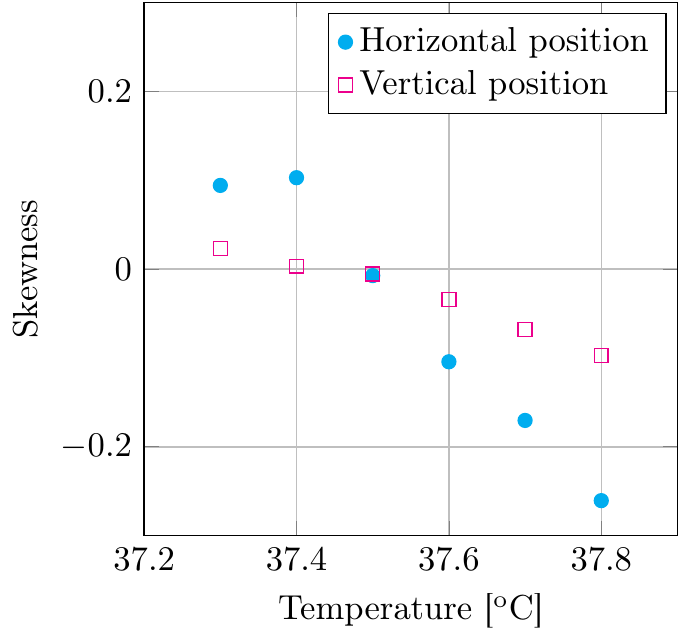}
\caption{Skewness of vertical and horizontal positions of the \emph{smart particle} versus the temperature of the cell center.}
\label{skew}
   \end{center}
   \end{figure}

To follow accurately the flow, the capsule and fluid density are carefully matched within 0.05\%. This is the main difficulty of the experiment. On figure \ref{skew} we plot the skewness of the distributions of the horizontal and vertical positions while varying mean temperature of the cell. Due to the flow symmetry, we assume that the particle {\bf ideally matches the fluid density}  when both horizontal and vertical skewness are close to 0, which {\bf happens}  for a temperature of 37.5\si{\degreeCelsius}. If the mean temperature is shifted of few tenth of degrees, symmetry is broken while the particle becomes less neutrally-buoyant. The effect is more dramatic on the horizontal position. To explain it, we assume a particle denser than the fluid. When traveling along the top plate, it is easily advected by a cold plumes: the particle goes downwards earlier during the travel. When traveling along the bottom plate, it is more difficult for a plume to advect the particle: it goes upwards only when reaching the corner of the cell. {\bf As a consequence}, the average horizontal particle trajectory is shifted close to the wall where hot plumes rise. This reasoning  {\bf holds also}  for a less dense particle and explain the large {\bf skewness of the  horizontal position}.

\begin{figure}
\begin{minipage}{0.49\textwidth}
\begin{center}
(a)
\includegraphics[width=1\textwidth]{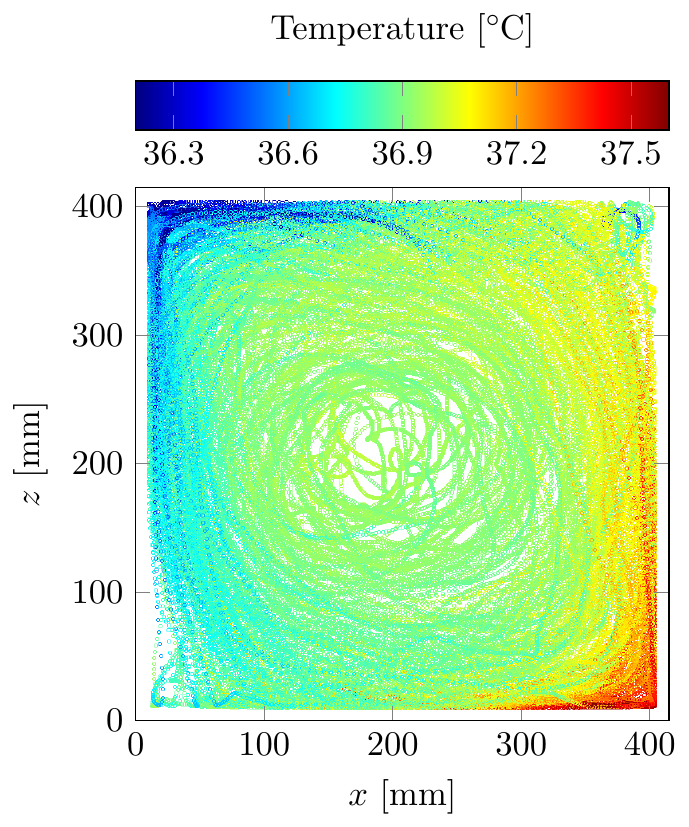}
\end{center}
\end{minipage}
\begin{minipage}{0.02\textwidth}
~~
\end{minipage}
\begin{minipage}{0.49\textwidth}
\begin{center}
(b)
\includegraphics[width=1\textwidth]{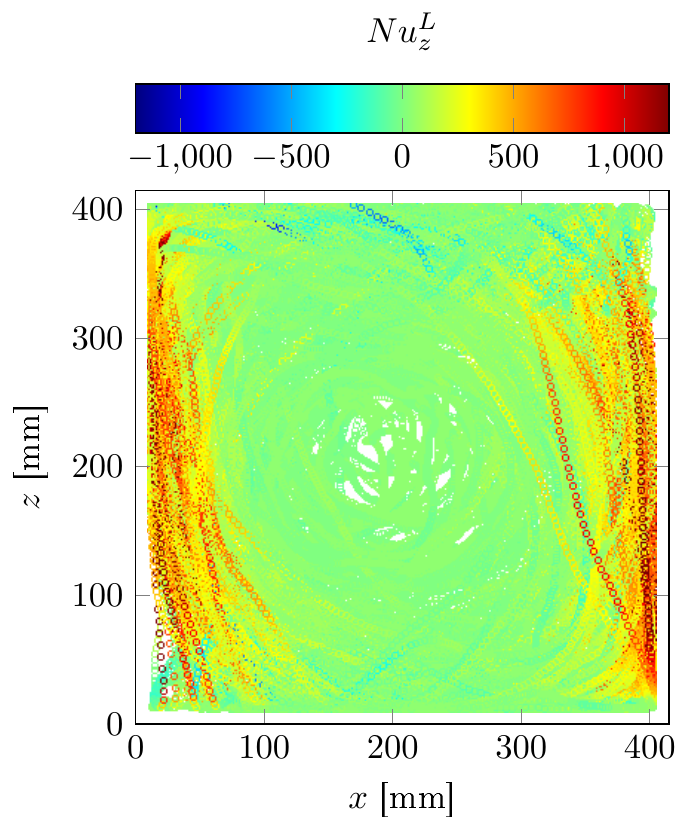}
\end{center}
   \end{minipage}
   \caption{(a) Flow temperature along the particle trajectory at $Ra=5.0\times 10^{10}$, six-hour acquisition. (b) Corresponding Lagrangian vertical Nusselt $Nu_z^L$. Trajectories are undersampled for visibility purposes.}
   \label{traj}
   \end{figure}

We performed measurements from 6 to 20.3 hours. Figure \ref{traj}(a) shows an example of {\bf a temperature measurement along the particle trajectory}. In this case, $Ra=5.0\times 10^{10}$ {\bf while the} acquisition time is six hours. Globally, the particle describes a loop in the counterclockwise direction (the rotation direction can change for other acquisitions). Its mean speed is 1 \si{\cm\per.\second}. Close to walls and plates, its speed is typically 2 to 3 \si{\cm\per.\second}.

The thermal boundary layer thickness can be computed as: 
\begin{equation}
\delta_{\theta} = \frac{H}{2Nu},
\end{equation} 

which corresponds to a thickness of less than 1 \si{mm} for the considered Rayleigh numbers, {\bf which is not directly accessible to our particle}. This is why we do not observe {\bf large temperature gradients  close to the top and bottom plates}. However, we {\bf clearly detect} hot and cold jets near the right and left walls, respectively. Temperature fluctuations in the \textcolor{red}{cell (outside the boundary layers)} are typically 1\si{\degreeCelsius} {\bf and are likely due to  advection by plumes}. Nevertheless, 
{\bf joint measurements of temperature and trajectory give indications that the particle movement is more influenced by the mean wind than by plumes}, except close to the vertical walls - inside hot and cold jets.

Knowing the velocity and temperature of the particle, we can compute the Lagrangian thermal flux. We use a normalized vertical Nusselt number \citep{ching, GL}:

\begin{equation}
Nu_z^L = 1+ \frac{H}{\kappa \Delta T} \left(T(t)-\langle T(t) \rangle_t \right) v_z(t),
\end{equation}

where $T(t)$ is the instantaneous temperature measured by the particle, $\langle T(t)\rangle_t$ is its mean along the trajectory and $v_z(t)$ is its vertical velocity. Figure \ref{traj}(b) shows the $Nu_z^L$ along the trajectory. Most of the vertical thermal transfer is symmetrically concentrated in the hot and cold jets, corresponding to plumes with $T(t)-\langle T(t) \rangle_t$ and $v_z(t)$ having the same sign. {\bf We also observe that} the vertical flux is highly asymmetric towards positive value up to 30 times the average vertical heat flux $\langle Nu_z^L\rangle_t=138$.

\section{Pseudo-Eulerian maps}

\subsection{Methodology and PIV measurements}

\begin{figure}
\begin{center}
\includegraphics[width=0.95\textwidth]{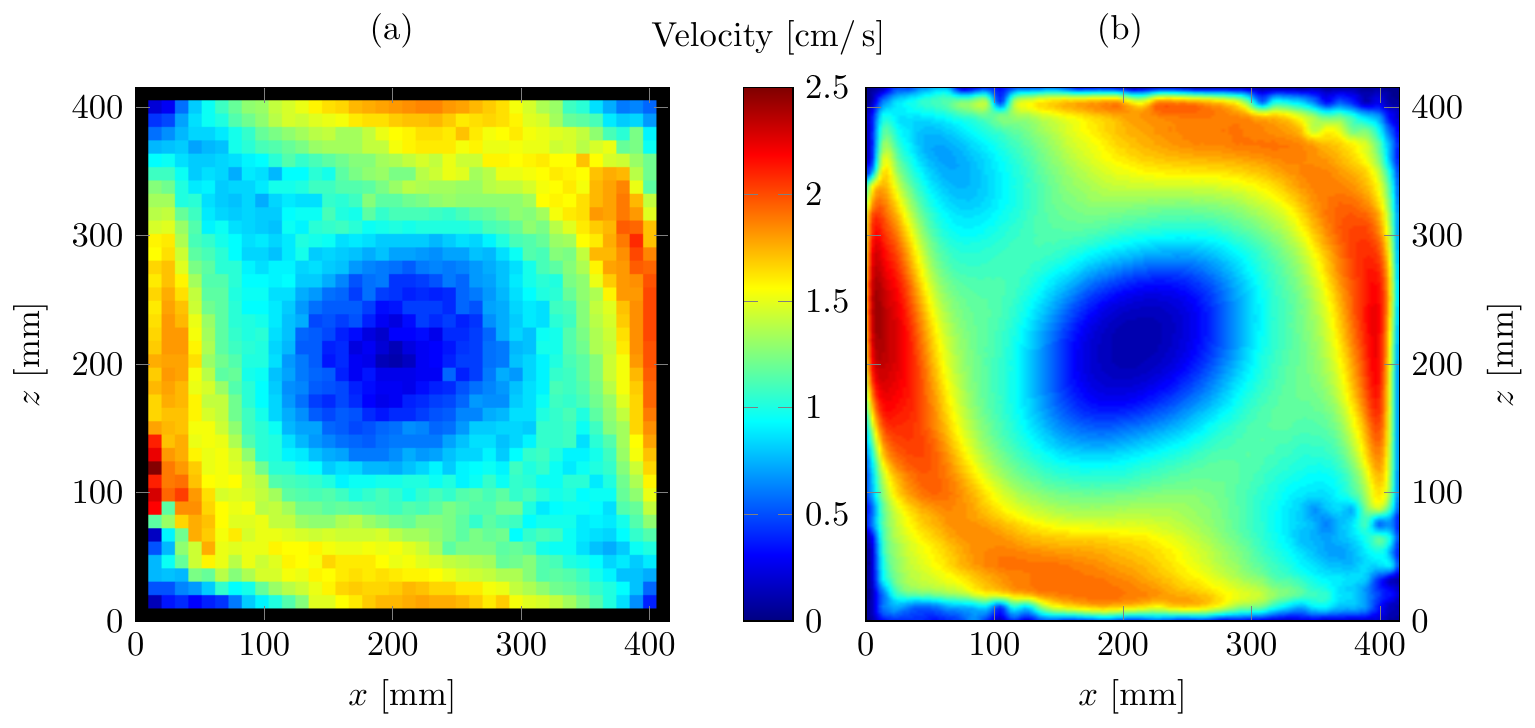}
\caption{(a) Pseudo-Eulerian module velocity field from Lagrangian data, $Ra=5.0 \times 10^{10}$. (b) Module velocity field obtained by PIV, $Ra=5.6 \times 10^{10}$.}
\label{PIV}
\end{center}
\end{figure}

{
The velocity distribution in the central region of the convection cell is close to that of a solid body rotation, {\bf a situation that hinders the particle}  from exploring easily this region. 
To \textcolor{blue}{obtain} a  correct resolution of the central region we performed very long experiments.
}
With more than twenty-hour measurement, we have enough data in the whole cell (including the central zone) to compute pseudo-Eulerian maps of several quantities. {\bf To do this}, we divide our cell in 1.04 \si{cm} $\times$ 1.04 \si{cm} squares {\bf and we compute} the average of the considered quantity in each \textcolor{red}{cell}. 
The accuracy of the method has been evaluated {\bf through} comparison \textcolor{red}{of the pseudo-Eulerian velocity field (figure \ref{PIV}(a))} with {\bf the results obtained by} Particle Image Velocimetry (PIV) measurements in the same cell (figure \ref{PIV}(b)) at similar Rayleigh numbers.
We performed PIV measurements with a 1.2 W, Nd:YVO$_4$ laser. Flow was seeded with Spherical 110P8 glass beads of 1.10$\pm$0.05 in density and of 12 \si{$\micro$m} average diameter. Twelve-hour acquisitions with one picture pair (frequency acquisition 20 Hz) every ten seconds were used to compute mean velocity fields. We used CIVx \citep{fincham} free software for analysis. Several passes are applied to picture pairs. For the first one, we cut out pictures in 30$\times$30 pixels$^2$ elementary boxes with 50$\%$ overlap. Search zones was one and a half larger.

 The velocity module obtained from PIV is in good agreement with observations from \cite{xia} for a similar cell. The velocity measured with the \emph{smart particle} is slightly lower than {\bf that obtained} with the PIV technique, probably due to several effects: a {\bf small difference in the}  Rayleigh number, particle size and inertia, and fluid slip on the particle surface. 

\subsection{Temperature and thermal flux maps}
\begin{figure}
\begin{minipage}{0.49\textwidth}
\begin{center}
(a)
\includegraphics[width=1\textwidth]{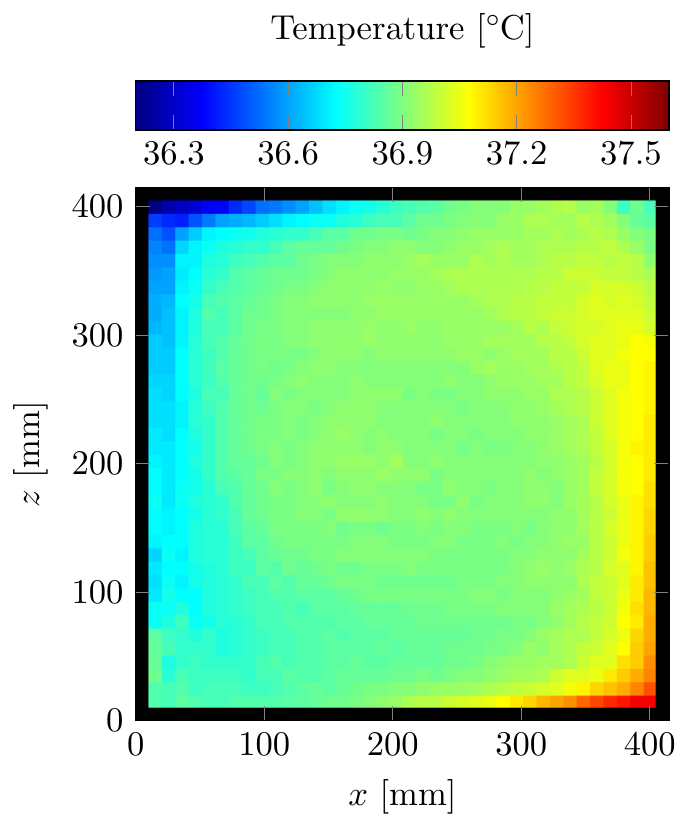}
\end{center}
\end{minipage}
\begin{minipage}{0.02\textwidth}
~~
\end{minipage}
\begin{minipage}{0.49\textwidth}
\begin{center}
(b)
\includegraphics[width=1\textwidth]{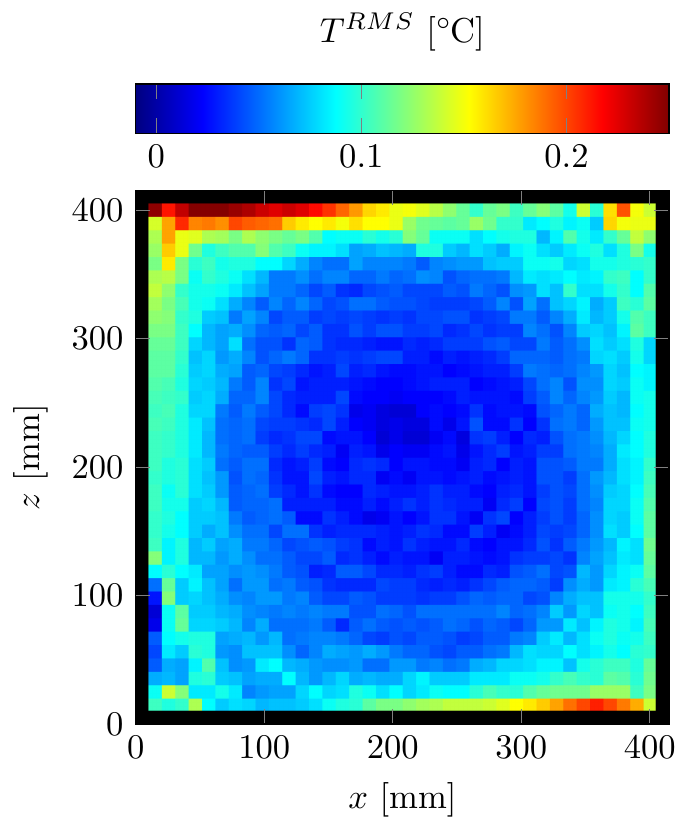}
\end{center}
   \end{minipage}
   \caption{(a) Pseudo-Eulerian temperature field and  (b) temperature quadratic fluctuations field. Measurements at $Ra=5.0 \times 10^{10}$.}
\label{map}
   \end{figure}

{
Our Lagrangian  method can be efficiently used to obtain pseudo-Eulerian map of temperature in the whole cell. 
To our knowledge, this is the first Lagrangian experiment giving the whole temperature map for this range of Rayleigh and Prandtl numbers. 
The pseudo-Eulerian temperature field is plotted in figure \ref{map}(a).
We observe that the flow is homogeneous in the bulk, whereas hot and cold jets dominate the regions close to the walls. 
Deviations of temperature from the bulk value are in particular seen in the top-left and bottom-right corners and are likely due to the effect of buoyant plumes 
driving the \emph{smart particle} along the vertical walls.
{\bf As suggested by \cite{scagliarini2014}, the mean wind acts stabilizing the boundary layers and reducing the plume emission activity.}
Moreover, the intense mixing 
makes the plume temperature close to the bulk temperature along the plates but in the top-left and bottom-right corners (where the mean wind is blocked and mixing cannot be observed). 
}
{\bf This situation is well represented in figure \ref{map}(b), where we observe large fluctuations in the corners but fewer fluctuations in the vertical jets and close to the plates.
 The observed slight asymmetry can be explained by a non-perfect particle-to-fluid density matching.} 
With velocity and temperature joint measurements, this is the only Lagrangian experimental technique that gives a pseudo-Eulerian thermal flux map. {\bf We observed that  $Nu_z^L$  is large inside both cold and hot jets, where vertical velocity $v_z$ is large, and indicates that the spatial distribution of $Nu_z^L$ is chiefly influenced by plumes}. 


   \begin{figure}
\begin{center}
\includegraphics[width=0.65\textwidth]{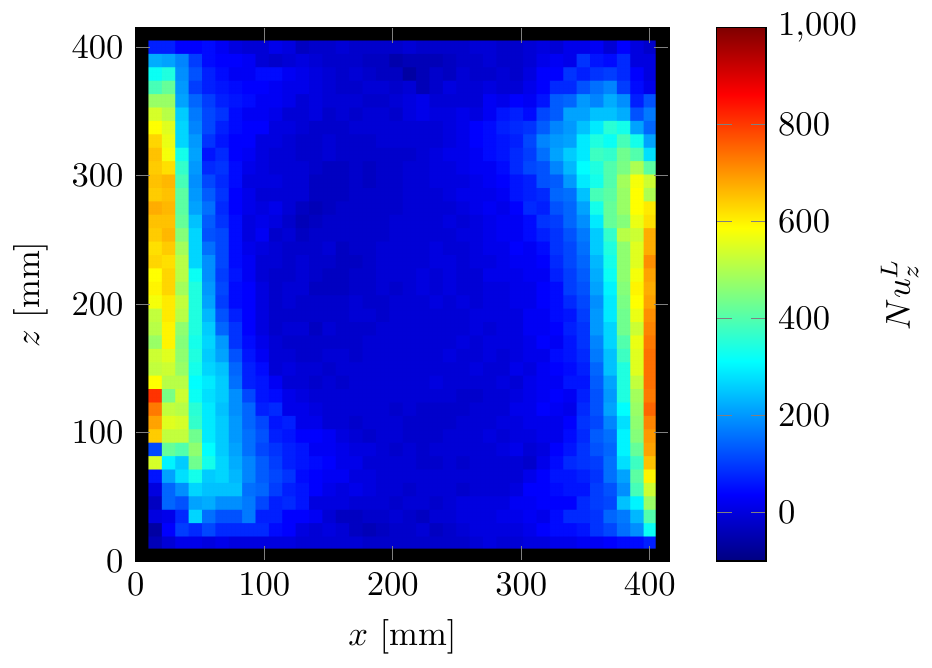}
\caption{Vertical Nusselt $Nu_z^L$ pseudo-Eulerian map at $Ra=5.0 \times 10^{10}$.}
\label{Nulzmap}
\end{center}
\end{figure}

\subsection{Velocity fluctuations}

{
Now we discuss velocity root mean square (RMS) maps obtained by our \emph{smart particle} in connection with the results obtained by PIV measurements.
}
From the velocity pseudo-Eulerian maps, we can compute smooth velocity mean fields by interpolation. This velocity value at each point $(x,z)$ is indicated as $v^E_i(x,z)$ where $i=x,z$ is the horizontal or vertical velocity, respectively. Thus, in each square $s$ described above, for all $(x_s,y_s)$ Lagrangian coordinates of the particle trajectory included in the square we have:

\begin{equation}
v_{i,s}^{RMS}=\sqrt{\langle\left( v^L_i(x_s,z_s)- v^E_i(x_s,z_s)\right)^2 \rangle_s},
\end{equation}

where $v^L_i(x_s,z_s)$ represents the Lagrangian velocity events inside the considered square and $\langle . \rangle_s$ is the average of these events. 
This is similar to the velocity RMS computed from PIV measurements for all $(x,y)$ in the cell: 

\begin{equation}
v_{i}^{RMS}(x,z)=\sqrt{\langle\left( v_i(x,z,t)-\langle v_i(x,z,t) \rangle_t\right)^2 \rangle_t}.
\end{equation}

\begin{figure}
\begin{center}
\includegraphics[width=1\textwidth]{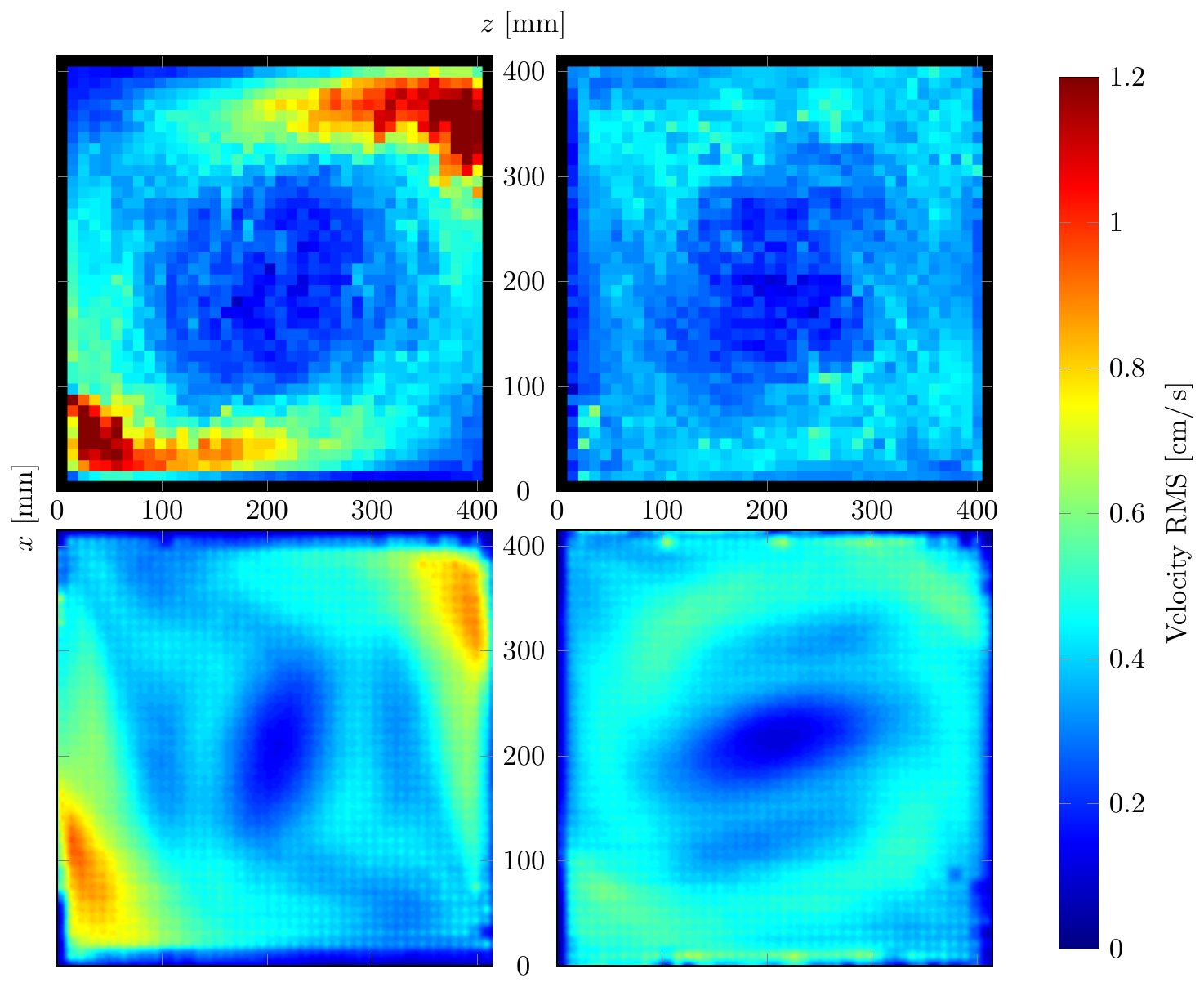}
\caption{Top-left: vertical velocity RMS pseudo-Eulerian map; top-right: horizontal velocity RMS pseudo-eulerian map. Bottom-left: vertical velocity RMS obtained by PIV; bottom-right: horizontal velocity RMS obtained by PIV. $Ra=5.0\times 10^{10}$ for pseudo-Eulerian maps and $Ra=5.6\times 10^{10}$ for PIV fields.}
\label{Vfluct}
   \end{center}
   \end{figure}

In these two definitions, both $v^E_i(x,z)$ and $\langle v_i(x,z,t) \rangle_t$ represent the mean flow. 
Figure \ref{Vfluct} compares vertical (left) and horizontal (right) velocity fluctuations from the \emph{smart particle} (top) and from PIV (bottom). 
{
First, we note that RMS values recorded by the particle are {\bf slightly} smaller compared to those evaluated by PIV.
This is mainly due to the filtering effect played by the particle on small scales fluctuations.
There is also a secondary effect due to {\bf a small difference in the } value of $Ra$ ($Ra=5\cdot 10^{10}$ for the Lagrangian particle and $Ra=5.6\cdot 10^{10}$ for PIV measurements).
The two regions characterized by the largest fluctuations are those where vertical plumes impinge on the horizontal walls and induce large turbulence patches. 
In our Lagrangian measurements, we also observe significant fluctuations along the horizontal plates: 
these fluctuations, which are not visible in the PIV measurements, are due to particle rebounds on the horizontal walls.
When the particle hits the wall, a positive vertical velocity becomes a negative vertical velocity and \textcolor{blue}{\emph{vice versa}}.
Differently, when the fluid approaches the wall, \textcolor{red}{a large part of} its vertical velocity is converted into horizontal one.
%
%
}
Consequently, for each rebound, we can suppose that the contribution of the particle to the RMS is twice {\bf that of the fluid}. This leads to a 50\% increase of the RMS value, which corresponds to the increase observed between pseudo-Eulerian and Eulerian vertical velocity RMS maps.

\section{Flow statistics}

\subsection{Probability density functions}

\begin{figure}
\begin{minipage}{0.49\textwidth}
\begin{center}
(a)
\includegraphics[width=1\textwidth]{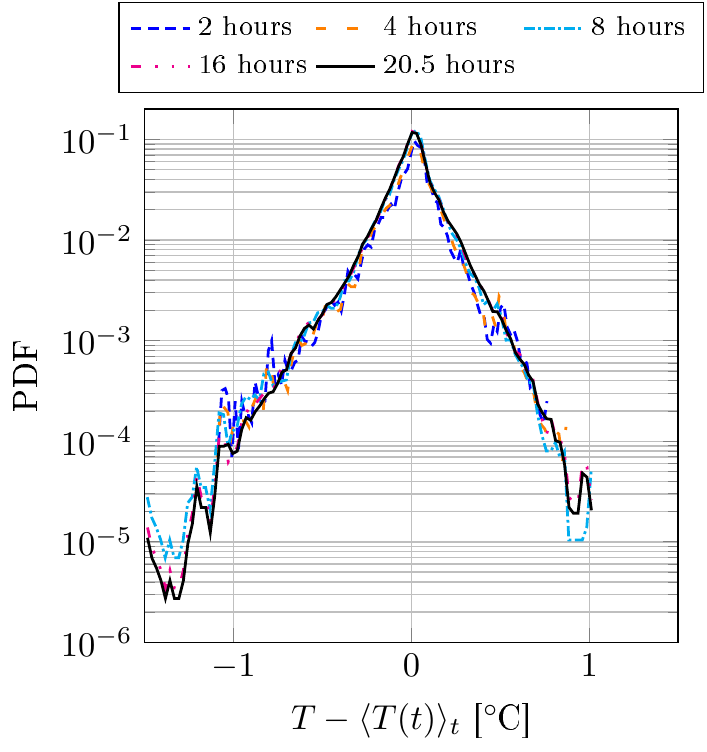}
\end{center}
\end{minipage}
\begin{minipage}{0.02\textwidth}
~~
\end{minipage}
\begin{minipage}{0.49\textwidth}
\begin{center}
(b)
\includegraphics[width=1\textwidth]{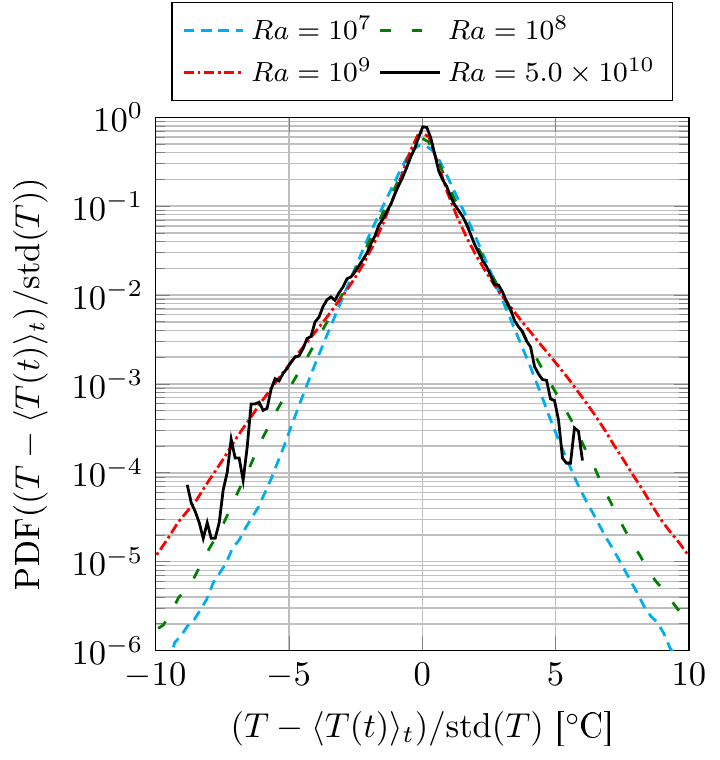}
\end{center}
   \end{minipage}
   \caption{(a) Probability Density Function (PDF) of the temperature fluctuations $T-\langle T(t)\rangle_t$ at $Ra=5.0\times 10^{10}$ for different measurement times. (b) Comparison between experimental and numerical results of the PDF of temperature fluctuations, $T-\langle T(t)\rangle_t$, normalized by the corresponding standard deviation. Solid line represents experimental data whereas dashed lines correspond to DNS data at different $Ra$.}
\label{PDF}
   \end{figure}

{
In figure \ref{PDF}(a) we show the probability density function (PDF) of the temperature fluctuations recorded by our instrumented particle at $Ra=5 \cdot 10^{10}$ for different measurement time ($t=2,4,8,16,20.5~hr$).
We note that the shape of the PDF becomes increasingly smooth for increasing measurement times.
Present results based on 20.5-hour measurements substantially improve previous results of \cite{gasteuil} obtained with a 2-hour acquisition. \textbf{Nevertheless, the global shape is conserved, and confirms the overall quality of the results by \cite{gasteuil}.}
The symmetric shape of the PDF is a further signature of the quality of the results (good buoyancy neutrality).
The PDF has sharp exponential tails, in agreement with previous Eulerian measurements \citep{belmonte} performed far from the boundary layers and using air at $Ra=4.8\times10^7$.
However, tails are wider due to the passage of the particle in hot and cold jets. This suggests that Eulerian measurements in Rayleigh-B\'enard convection are delicate: since the flow is largely inhomogeneous, the position at which the measurement is taken is fundamental. 
The generalization of the behaviour of the entire cell based on a local measurement requires a lot of care.
In Fig. \ref{PDF}(b), we compare experimental results of the PDF of temperature fluctuations obtained in the 20.5-hour measurement with our results from numerical simulations at different Rayleigh numbers.
Results are normalized by the corresponding temperature standard deviation, std$(T)$. 
Even though numerical experiments are carried out at smaller Rayleigh number, the agreement between experiments and simulations is satisfactory
when $Ra>10^8$.
Deviations between experimental and numerical results are observed only for $(T-\langle T(t)\rangle_t)/std(T)>3$, hence highlighting the accuracy of the present experiments.
Indeed, our numerical simulations can be seen as an ideal experiment, since we sample the flow-field with massless (pointwise) fluid tracers (no size/inertia effect of the Lagrangian probes).
Within this framework, the difference between experiments and simulations can give indications on the effect of the size of our \emph{smart particle}:
due to the finite-size, our \emph{smart particle} acts as a filter for small/short space/time scale events, and numerical experiments show larger tails.
}

\begin{figure}
\begin{minipage}{0.49\textwidth}
\begin{center}
(a)
\includegraphics[width=1\textwidth]{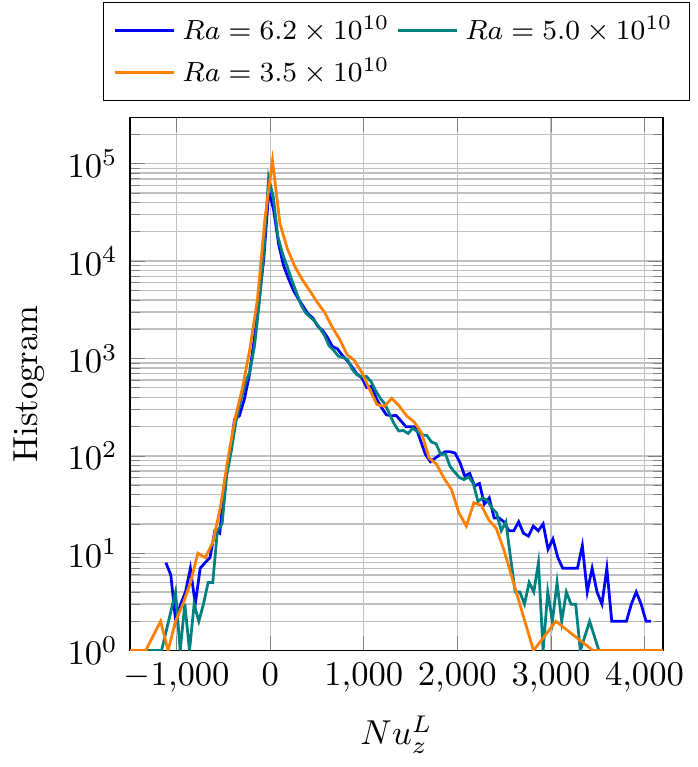}
\end{center}
\end{minipage}
\begin{minipage}{0.02\textwidth}
~~
\end{minipage}
\begin{minipage}{0.49\textwidth}
\begin{center}
(b)
\includegraphics[width=1\textwidth]{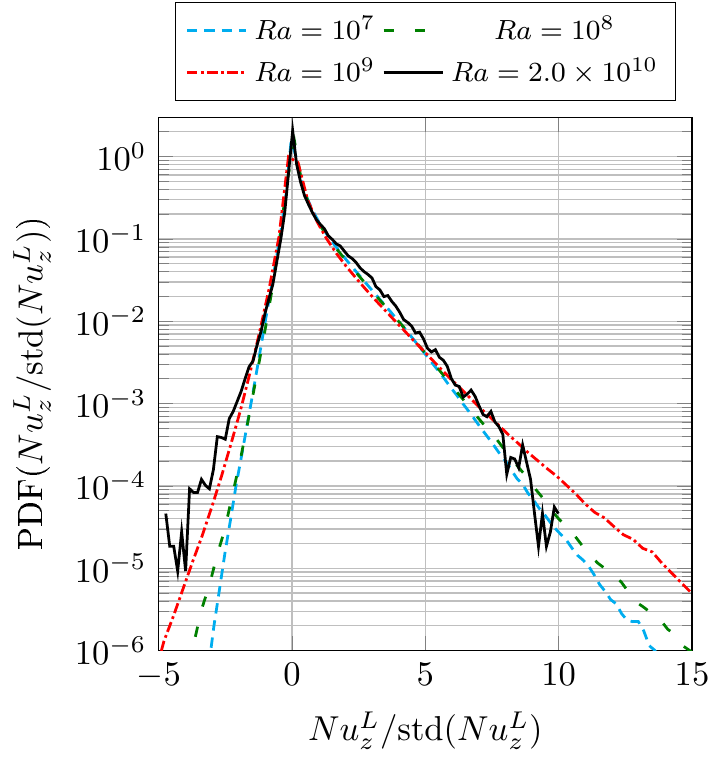}
\end{center}
   \end{minipage}
   \caption{(a) PDF of the vertical thermal flux $Nu_z^L$ for different Rayleigh numbers, six-hour measurements. (b) Comparison of the vertical Nusselt PDF $Nu_z^L$ normalized by the standard deviation. Solid line represents experimental data whereas dashed lines correspond to DNS ones.}
\label{PDFNu}
   \end{figure}

{
As previously mentioned, with our instrumented particle we are able to record simultaneously velocity and temperature, hence we can compute the local value of the vertical Nusselt number $Nu^L_z$.
Figure \ref{PDFNu}(a) shows the \textcolor{red}{histogram of the} vertical Nusselt for three different Rayleigh numbers. 
The most probable value is $Nu^L_z=0$, corresponding to the time during which the particle has a horizontal trajectory or is advected by the mean wind far from the walls  (hence far from plumes), where no significant vertical heat flux is observed. 
Interestingly, there is a larger positive tail, which is the signature of near wall intense events  \citep{gasteuil}. 
When $Ra$ increases, the  shape of the {\bf histogram} does not change, but we observe more intense events at higher $Ra$. 
In figure  \ref{PDFNu}(b) we compare our experimental and numerical results of $PDF(Nu^L_z)$. Results are normalized by their corresponding standard deviation, $std(Nu^L_z)$. We observe a very good agreement between experiments and simulations over the entire range of measured $Nu^L_z/std(Nu^L_z)$.
Deviations are seen only for extreme and rare events, $Nu^L_z/std(Nu^L_z)>7$, due to the filtering effect of the \emph{smart particle} size on velocity/temperature fluctuations.
}

\subsection{Spectral analysis}

\begin{figure}
\begin{minipage}{0.49\textwidth}
\begin{center}
(a)
\includegraphics[width=1\textwidth]{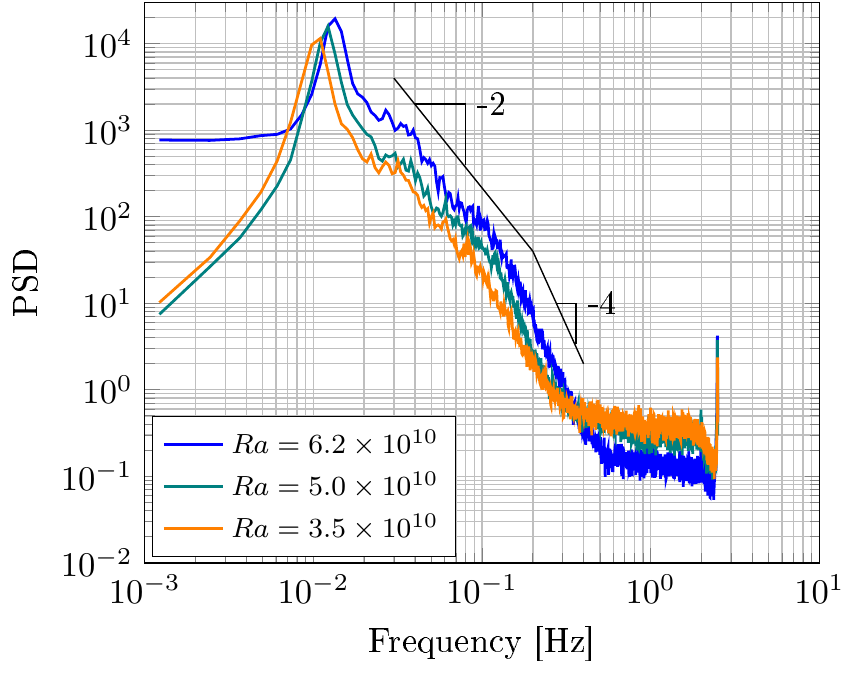}
\end{center}
\end{minipage}
\begin{minipage}{0.02\textwidth}
~~
\end{minipage}
\begin{minipage}{0.49\textwidth}
\begin{center}
(b)
\includegraphics[width=1\textwidth]{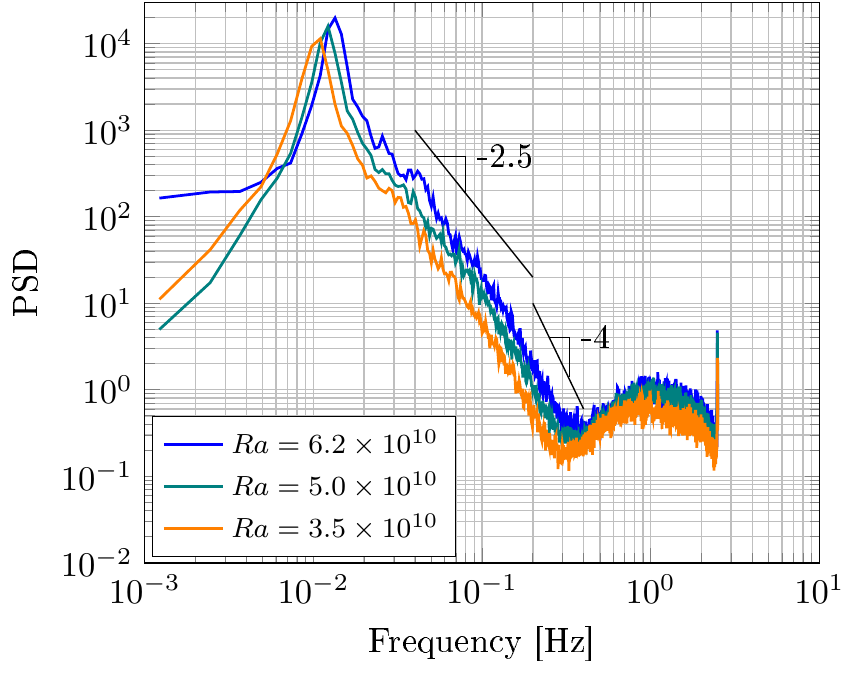}
\end{center}
   \end{minipage}
   \caption{(a) Vertical and (b) horizontal velocity spectra at different Rayleigh numbers.}
\label{spectresv}
   \end{figure}

\begin{figure}
\begin{center}
\includegraphics[width=0.55\textwidth]{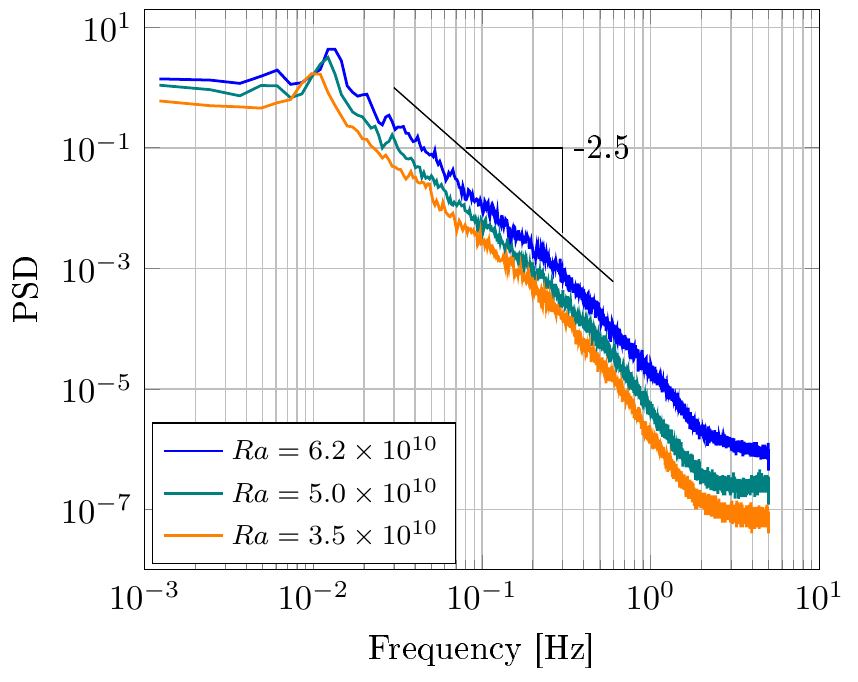}
\caption{Temperature spectra at different Rayleigh numbers.}
\label{spectreT}
   \end{center}
   \end{figure}

{
We conclude our discussion with 
a spectral analysis of velocity and temperature Lagrangian time series recorded at different Rayleigh numbers. 
}
Figures \ref{spectresv}(a), \ref{spectresv} (b) and \ref{spectreT} show the frequency spectra (for six-hour acquisitions) of vertical velocity, horizontal velocity and temperature, respectively. We observe three distinctive characteristics. First, a peak appears {\bf at a frequency} $f\simeq 1.25\times 10^{-2}$ Hz (80 s): it is consistent with the typical {\bf large scale }turnover time. 
{
Yet, when $Ra$ increases, the spectrum shifts upwards, indicating a more intense dynamics.
Finally, a cut-off occurs at 0.15 Hz for velocity and 0.5 Hz for temperature. This latter observation indicates a finite-size particle effect, which is less important for temperature owing to the thermistors lower size.
}

From the proposed scaling laws (lines in figures \ref{spectresv}-\ref{spectreT}) we note some interesting features.
 A $f^{-2}$ power law characterizes the vertical velocity, whereas a $f^{-2.5}$ power law characterizes the horizontal velocity. 
{\bf A similar } flow anisotropy has been also observed by \cite{qureshi} in a Von-K\'arm\'an flow.
{
From the above observations, we can infer the following physical interpretation.
The particle has a large vertical velocity when it enters hot and cold jets along the lateral walls, 
whereas it has a large horizontal velocity when it is driven by the mean wind along the top and bottom walls.
Since  the hot and cold jets are characterized by more intense fluctuations, the corresponding spectra have a weaker slope.
}
%
After the cut-off, a classical $f^{-4}$ power law appears for both spectra. 
{
Concerning the temperature spectra, we observe a $f^{-2.5}$ power law. 
Note that \textcolor{red}{due to the anisotropic and inhomogeneous flow condition we do not find a $f^{-2}$ slope}. Yet, temperature can not be considered as a pure passive scalar, in particular inside hot and cold jets.  Unfortunately, at present we do not have a quantitative explanation for this power law.
Results from numerical simulations (not shown here) differ from the present experimental results and essentially follow the classical K41 scaling.
This difference is mainly due to the different flow configuration: our simulations are run in a domain with top and bottom walls but no side walls, 
whereas experiments are run in a square enclosure. 
}

\section{Conclusion}
{
   In this work, we have used an improved version of our \emph{smart particle} to perform new experimental measurements in  Rayleigh-B\'enard convection at different Rayleigh numbers (up to $Ra=6.2 \cdot 10^{10}$). 
To corroborate the experiments and 
   isolate possible finite-size effects, we have also carried out  Direct Numerical Simulations of Rayleigh-B\'enard turbulence with Lagrangian tracking of massless tracers.
 
In the experiments, due to the extended autonomy of the particle, we are able to sample velocity and temperature at the particle position for long periods of time, up to twenty hours. 
 This long data recording \textcolor{blue}{allows} not only to acquire long time-series describing  the temporal evolution of a turbulent flow,
 but also to build pseudo-Eulerian maps of the flow field and to compute converged PDF and spectra. 
The particle trajectory is driven by the interaction between the mean large scale circulation (along the horizontal walls) and the thermal plumes generating vertical hot and cold jets.
Velocity and temperature fluctuations are essentially concentrated in the bottom-right and top-left corners, and denotes strong turbulence events in these regions.

Upon comparison between experimental and numerical results, we are able to demonstrate the accuracy of our experimental technique in recovering all the fundamental statistical features of the flow. 
We finally computed velocity and temperature frequency spectra.
Interestingly, horizontal and vertical velocity spectra exhibit  different scaling ($f^{-2}$ and $f^{-2.5}$, respectively), 
as a consequence of the strong flow anisotropy in the vertical and horizontal directions.
For temperature, we observe a steep $f^{-2.5}$ scaling, likely representing the hybrid passive and active scalar behavior of temperature.
}

  A further study of these observations will be the subject of a future paper. As the flow is dominated by a mean vortical structure, we can remove it from the recorded signal to study the behavior of turbulence fluctuations and the  interactions between fluctuations and mean structure along the line proposed by \cite{machicoane2}  for Von-K\'arm\'an flows.
   
   \vspace*{0.5cm}
   
   \textbf{Acknowledgements:} We thank Marius Tanase for his precious technical help about electronic device of the particle, Denis Le Tourneau and Marc Moulin for the manufacture of the cell. PIV measurements were made possible with the help of PSMN computing resources. Many thanks to Romain Volk and Micka\"el Bourgoin for fruitful discussions.
   
      \bibliographystyle{jfm}

\bibliography{biblio_Joe}

\end{document}